\documentclass[letterpaper, 10 pt, conference]{ieeeconf}

\IEEEoverridecommandlockouts   

\usepackage[dvips]{graphicx}
\usepackage{amsmath}
\usepackage{amssymb}

\DeclareMathOperator*{\argmin}{argmin}

\title{\LARGE \bf
Experimental Implementation of an Invariant Extended Kalman Filter-based Scan Matching SLAM}

\author{Martin Barczyk, Silv{\`e}re Bonnabel, Jean-Emmanuel Deschaud and Fran{\c c}ois Goulette%
\thanks{The authors are with the Centre de Robotique, Unit{\'e} Math{\'e}matiques et Syst{\`e}mes, Mines ParisTech, 75272 Paris Cedex 06, France
        {\tt\small \{firstname.lastname\}@mines-paristech.fr} }%
}

\begin{document}

\maketitle
\thispagestyle{empty}
\pagestyle{empty}

\begin{abstract}

We describe an application of the Invariant Extended Kalman Filter (IEKF) design methodology to the scan matching SLAM problem. We review the
theoretical
foundations of the IEKF and its practical interest of guaranteeing robustness to poor state estimates, then implement the filter on a wheeled robot hardware
platform.
The proposed design is successfully validated in experimental testing.
\end{abstract}

\section{Introduction}
\label{sec:intro}

Simultaneous Localization and Mapping or SLAM is an active area of research in robotics due to its use in emerging applications such as
autonomous driving and piloting, search-and-rescue missions and mobile cartography~\cite{AANGL04}. SLAM is fundamentally a sensor fusion problem, and as such
it is typically handled via an Extended Kalman Filter (EKF), although a number of direct nonlinear designs have also been
proposed e.g.~\cite{LTM10,VCSO10,Hua10,SM11,GFJS12}.

Applying an EKF to the SLAM problem was first seen in~\cite{SSC86}, using a state vector which keeps track of landmark positions and whose
dimension grows as new landmarks enter into view. Using this technique for experimentally validated localization and
mapping was seen in e.g.~\cite{DNCDC01}. An alternative non-EKF-based approach to SLAM was first proposed in~\cite{LM94} relying on matching successive
scans of
the environment from onboard sensors in order to localize the vehicle and construct a map of the environment. This second approach was experimentally
implemented in
e.g.~\cite{Nuc04} for autonomously mapping an abandoned mine. The scan matching-based method uses vehicle odometry to obtain estimates of the 
vehicle pose by numerical integration of the dynamics, known as dead-reckoning~\cite{Farrell}. The estimated poses are then used to localize new scans
before
matching them with previous scan(s). This procedure can be seen as a sensor fusion problem between noisy odometry and scan matching data, and thus handled by
an EKF as proposed in~\cite{NBN06}. We will employ the EKF-based scan matching approach to SLAM throughout the rest of this paper.

The EKF works by linearizing a system about its estimated trajectory, then using estimates obtained from an observer for this linearized model
to correct the state
of the original system. In this way the EKF relies on a closed loop which can be destabilized by sufficiently poor estimates of the trajectory, known
as divergence~\cite{Brown}. Clearly, reducing or eliminating the dependence of the EKF on the system's trajectory would increase the
robustness of the
overall system. An emerging methodology to accomplish this goal is the Invariant EKF~\cite{Bon07,BMS09}, built on the theoretical foundations of invariant
(symmetry-preserving) observers~\cite{BMR08}, specialized to the case of Lie groups in~\cite{BMR08a,BMR09}. The IEKF technique has already
demonstrated experimental performance improvements over a conventional
EKF in aided inertial navigation designs~\cite{MS10,BL13}. Applying the IEKF to scan matching SLAM was first demonstrated in~\cite{HBG12}. The
contribution of
the present paper is to design and implement an IEKF-based scan matching SLAM for a wheeled indoor robot and experimentally validate the design.

\section{Mathematical Preliminaries}

\subsection{System Dynamics}

The dynamics of a 6 DoF vehicle are governed by
\begin{equation}
\begin{aligned}
\dot{R}&=R S(\omega)\\
\dot{p}&= R \mu
\end{aligned}
\label{eq:dyncomponents}
\end{equation}
where $R\in SO(3)$ is a rotation matrix measuring attitude of the vehicle, $\omega
\in
\mathbb{R}^3$ is the body-frame angular
velocity vector, $S(\cdot)$ is the $3\times 3$ skew-symmetric matrix such that $S(x)y=x \times y$, the $\mathbb{R}^3$ cross-product, $p\in\mathbb{R}^3$ is the
position vector of the vehicle expressed in coordinates of the ground-fixed frame,
and $\mu$ is the velocity vector of the vehicle expressed in body-frame coordinates. We assume $\omega$ and $\mu$ are directly measurable using
on-board sensors, e.g.~via a triaxial rate gyro and odometers on a wheeled vehicle. The vehicle state space can be identified as
the Special Euclidean group $SE(3)=SO(3)\times \mathbb{R}^3$, where each $X\in SE(3)$ is written as the $4\times 4$ homogeneous matrix~\cite{MLS94}
\begin{equation*}
X=\begin{bmatrix}
R & p \\
0 & 1
\end{bmatrix}
\end{equation*}
In this way dynamics~\eqref{eq:dyncomponents} can be compactly rewritten as
\begin{equation}
\dot{X}=X \Omega\quad\text{where}\quad \Omega=\begin{bmatrix}
S(\omega) & \mu \\
0 & 0
\end{bmatrix}
\label{eq:dynSE3}
\end{equation}
We assume that the pose $X$ of the vehicle can be obtained by performing the
procedure of scan matching described in Section~\ref{sec:scanmatch}, such that the output of the system is
\begin{equation}
Y=X
\label{eq:outSE3nonoise}
\end{equation}

\subsection{Geometry of $SE(3)$}
\label{sec:SE3geom}

The system dynamics~\eqref{eq:dynSE3} evolve on the Lie group $SE(3)$, which provides access to a
number of useful results from differential geometry.

For any Lie group $G$, the exponential map $\text{exp}:\text{Lie}(G) \to G$ maps elements of the Lie algebra
to the Lie group, where $\text{Lie}(G)\cong T_e G$, the tangent space to $G$ at the identity element $e$. The following properties of this map will be used in
the sequel:
\begin{itemize}
\item $\text{exp}$ is a smooth map from $\text{Lie}(G)$ to $G$
\item $\exp$ restricts to a diffeomorphism from some neighbourhood of $0$ in $\text{Lie}(G)$ to a neighbourhood of $e$ in $G$
\item $(\text{exp}\, X)^{-1}=\text{exp}(-X)$
\end{itemize}
For the case of $G=SE(3)$ written in homogeneous matrix coordinates, $\text{Lie}(SE(3)) \cong se(3)$ where the Lie algebra members $\xi \in se(3)$ are written
as~\cite{MLS94}
\begin{equation}
\xi = \begin{bmatrix}
S(v_R) & v_p \\
0 & 0
\end{bmatrix}:=H\left(\begin{bmatrix}v_R \\ v_p \end{bmatrix}\right), \quad v_R,v_p \in \mathbb{R}^3
\label{eq:se3algebraform}
\end{equation}
and $\text{exp}:se(3) \to SE(3)$ is the standard matrix
exponential
\begin{equation*}
e^\xi:=I+\xi+\frac{1}{2!}\xi^2+\cdots
\end{equation*}
The map $H:\mathbb{R}^6 \to se(3)$ defined in~\eqref{eq:se3algebraform} is a vector space isomorphism. 
Remark that the $\Omega$ term in~\eqref{eq:dynSE3} can be written as $\Omega=H([\omega,\mu])$.

As stated above, $\text{exp}$ restricts to a diffeomorphism from a neighborhood of $0$ in $\text{Lie}(G)=se(3)$ to a neighborhood $U$ of $e$ in $G=SE(3)$. The
inverse of $\text{exp}$ is the logarithmic map $\text{log}:G \to \text{Lie}(G)$ such that $\text{log}\circ \text{exp}(g) =g,\,\forall g\in U$. The formula for
$\text{log}: SE(3) \to se(3)$ using homogeneous matrices is~\cite{MLS94}
\begin{equation}
\begin{aligned}
\text{log}& \begin{bmatrix}
R & p \\ 0 & 1
\end{bmatrix} = \begin{bmatrix}
S(a) & A^{-1} p \\
0 & 0
\end{bmatrix}, \quad S(a)=\theta S(\omega)\\
  \theta&= \text{acos}\left(\frac{\text{trace}\, R-1}{2}\right),\quad S(\omega) = \frac{1}{2\sin\theta} (R-R^T) \\
A&= I + \frac{S(a)}{\|a \|^2} (1-\cos \|a\|) + \frac{S(a)^2}{\|a \|^3}( \| a \| - \sin \|a\|)
\end{aligned}
\label{eq:SE3log}
\end{equation}

We now discuss approximations which will be employed in the sequel. Consider the case where $X=e^\xi$ is close to identity. We know the exponential map
restricts to a diffeomorphism from a neighborhood of zero in $se(3)$ to a neighborhood of identity in $SE(3)$, thus $\xi$ is close to the 
matrix $0_{4\times 4}$. This means $\xi^2$ and all subsequent
terms in $e^\xi$ can be dropped as higher-order terms, such that
\begin{equation}
X \approx I + \xi, \qquad  \text{$X\in SE(3)$ close to identity}
\label{eq:Xapprox}
\end{equation}
From $(\text{exp}\, X)^{-1}=\text{exp}(-X)$ we have
\begin{equation}
X^{-1} \approx I - \xi, \qquad  \text{$X\in SE(3)$ close to identity}
\label{eq:Xinvapprox}
\end{equation}
Still considering $X$ close to identity, in~\eqref{eq:SE3log} we have $\text{trace}\, R
\approx 3 \Longrightarrow \theta \approx 0 \Longrightarrow a \approx 0$, thus $\| a \|$ is small and so
\begin{equation*}
\begin{aligned}
S(a)&\approx \frac{\theta}{2 \theta} (R-R^T) = \frac{R-R^T}{2}\\
A&\approx I + \frac{S(a)}{\|a \|^2} (1-1) + \frac{S(a)^2}{\|a \|^3}( \| a \| - \|a\|) = I
\end{aligned}
\end{equation*}
This leads to the projection map $\pi:SE(3)\to se(3)$,
\begin{equation}
X=\begin{bmatrix}
R & p \\ 0 & 1
\end{bmatrix} \Longrightarrow \pi(X)=\begin{bmatrix}
\frac{R-R^T}{2} & p \\ 0 & 0
\end{bmatrix} 
\label{eq:SE3pi}
\end{equation}

The map $\pi$ is defined everywhere but it is the inverse of $\text{exp}$ only when $X\in SE(3)$ is close to identity. In this case with
$X=e^\xi$ we have $\pi(X)=\xi$ and by~\eqref{eq:Xapprox} we obtain
\begin{equation}
X - I \approx \pi(X), \qquad  \text{$X\in SE(3)$ close to identity}
\label{eq:XminusIapprox}
\end{equation}

\subsection{Iterative Closest Point algorithm}
\label{sec:ICP}

Assume the vehicle is equipped with sensors which capture 3-D scans of the environment, either directly using a time-of-flight LiDAR or
indirectly using image-based reconstruction e.g.~a Kinect. These scans are represented as sets of points expressed in
coordinates of the body-fixed frame, known as point clouds. The Iterative Closest Point (ICP) algorithm~\cite{BM92} is an iterative procedure to
find the $\delta X\in SE(3)$ transformation which aligns cloud $A:=\{a_i\}$, $1\leq i \leq N_A$ to cloud $B:=\{b_i\}$, $1\leq i \leq N_B$ by minimizing the
least-squares cost function
\begin{equation}
 \sum_{i=1}^{N_A} \| \delta X a_i - b_i' \|^2 \quad \text{where} \quad b_i' := \argmin_{b_i \in B} \| b_i - a_i \|^2
\label{eq:ICPcost}
 \end{equation}
Points $\{ b_i'\}$ can be found using an approximate nearest-neighbour search algorithm. The ICP algorithm guarantees monotonic
convergence to a local minimum of the cost function~\eqref{eq:ICPcost}. 

\subsection{Scan Matching algorithms}
\label{sec:scanmatch}

By construction the vehicle pose $X$ transforms elements of the 3-D point cloud $A=\{a_i\}$ from the robot-fixed frame to
the ground-fixed frame.

Because successive scans contain elements from the same scene, we can estimate the pose of the robot $\hat{X}$ by repeatedly using the ICP algorithm to
compute
the transformation $\delta X$ between successive point clouds, known as scan matching. The algorithm proceeds as follows:
\begin{enumerate}
 \item Define the initial robot pose as $\hat{X}=I$ and assign the initial point cloud to $A$
 \item Whenever a new point cloud is available,
 \begin{enumerate}
  \item Transfer contents of cloud $A$ to cloud $B$
  \item Assign the new point cloud to $A$
  \item Compute $\delta X$ aligning $A$ to $B$ via ICP 
  \item Let $\hat{X}=\hat{X} \delta X$
 \end{enumerate}
 \item Goto 2.
\end{enumerate}
There are two issues with this approach. First, if the scanning rate is not sufficiently fast, the ICP algorithm may compute an incorrect $\delta X$ due to
convergence to an alternative local minimum of~\eqref{eq:ICPcost}. A second problem is loss of observability: for certain types of environments e.g.~moving
along a long featureless corridor, successive point cloud scans will be identical such that $\delta X=I$, and so the estimated pose will not get updated despite
the
robot moving.

A more robust pose estimation scheme makes use of the on-board inertial sensors. We numerically integrate~\eqref{eq:dynSE3} to obtain an initial estimate of
the current pose $\hat{X}$, which is used to pre-align the scans in the body-fixed frame. The algorithm proceeds as follows:
\begin{enumerate}
 \item Assign the initial robot pose $\hat{X}=I$ and initial point cloud to $B$
 \item Numerically integrate $\dot{\hat{X}}=\hat{X} \Omega$ using the current sensor signals
 \item Whenever a new point cloud is available,
 \begin{enumerate}
 \item Assign the new cloud to $A$ and compute $\hat{X}^{-1}\{b_i\}$
  \item Compute $\delta X$ aligning $A$ to $\hat{X}^{-1}\{b_i\}$ via ICP 
  \item Update estimated pose as $\hat{X}= \hat{X} \delta X$
  \item Using this updated pose compute $\hat{X} \{a_i\}$ and store the result as $B$
\end{enumerate}
 \item Goto 2.
\end{enumerate}
In this odometry-aided scan matching algorithm cloud $\hat{X}^{-1}\{b_i\}$ is pre-aligned to $A$, reducing the risk of the ICP converging to an
incorrect local
minimum even at low scanning rates. Meanwhile in unobservable environments where the two clouds are indistinguishable, $\hat{X}$ will still get
updated at step
2.

\subsection{Computing covariance of pose estimates}
\label{sec:ICPerrormodel}

For the second algorithm in Section~\ref{sec:scanmatch}, the cost function to be
minimized by the ICP at every new scan is
\begin{equation}
 f(\delta X) = \sum \| \delta X a_i - \hat{X}^{-1} b_i \|^2
\label{eq:costICPb}
\end{equation}
and the corresponding pose estimate output is
\begin{equation}
 Y=X=\hat{X} \delta X
\label{eq:YoutputICPb}
\end{equation}
Assuming that the $\hat{X}$ obtained from odometry integration is reasonably accurate, the clouds $\{a_i\}$ and $\hat{X}^{-1}
\{b_i\}$ in~\eqref{eq:costICPb} will be pre-aligned such that $\delta X$ will be close to identity. We can thus employ~\eqref{eq:Xapprox}
in~\eqref{eq:costICPb}:
\begin{equation}
 f(\xi) = \sum \|  a_i - \hat{X}^{-1} b_i + \xi a_i \|^2
 \label{eq:costICPb3}
\end{equation}
Expanding $\xi a_i$ via~\eqref{eq:se3algebraform} gives
\begin{equation*}
\xi a_i = S(v_R) a_i + v_p =  \begin{bmatrix} -S(a_i) & I_3
\end{bmatrix}  \begin{bmatrix} v_R \\ v_p \end{bmatrix}
\end{equation*}
and by defining $B_i:=[S(a_i) \quad -I_3]$, $x:=[v_R \quad v_p]^T$ and $y_i :=  a_i - \hat{X}^{-1} b_i$ the cost function \eqref{eq:costICPb3} is rewritten as
\begin{equation}
f(x)=\sum \| y_i - B_i x \|^2
\label{eq:costICPb4}
\end{equation}
meaning that as long as the odometry is reasonably accurate, the ICP will behave as a linear least-squares estimator of $x$.

Because the ICP aligns successive scans, we first assume the residuals $y_i - B_i x$ will have mean zero and be normally distributed with
diagonal covariance
$\sigma^2 I_3$ representing additive Gaussian sensor noise on the point cloud. Under these conditions the covariance of the least-squares estimate $\hat{x}$
which minimizes~\eqref{eq:costICPb4} is given by~\cite{Kay} 
\begin{equation*}
\sigma^2 \left[ \sum B_i^T B_i \right]^{-1} = \sigma^2 \left[ \sum_{i=1}^N \begin{bmatrix} -S(a_i)^2 & S(a_i) \\ -S(a_i) &  I_3
\end{bmatrix} \right]^{-1}
\end{equation*}
where $a_i$ are points of the current cloud in the robot-fixed frame. However the assumption of independent Gaussian noises is unrealistic, e.g.~indicating
that sub-millimeter accuracy can be achieved by scan matching two point clouds obtained from a Kinect for which $\sigma\approx 5\text{ cm}$. We thus
propose as a relevant approximation to rescale the above covariance matrix as
\begin{equation}
N \sigma^2 \left[ \sum_{i=1}^N \begin{bmatrix} -S(a_i)^2 & S(a_i) \\ -S(a_i) &  I_3
\end{bmatrix} \right]^{-1}:=C
 \label{eq:covICPb}
\end{equation}
We then have $\hat{x}\sim \mathcal{N}(x,C) \Longrightarrow \hat{x}=x+\nu$ where
$\hat{x}=[\hat{v}_R \quad
\hat{v}_p ]^T$ and $\nu:=[\nu_R \quad \nu_p]^T\sim \mathcal{N}(0,C)$.

Returning to the pose output~\eqref{eq:YoutputICPb} and employing~\eqref{eq:Xapprox} with $\hat{\xi}:=H(\hat{x})$ via~\eqref{eq:se3algebraform} denoting the
estimate produced by the ICP from measured (noisy) data, the corresponding estimated pose output is
\begin{equation*}
 Y_m= \hat{X} (I+\hat{\xi})
\end{equation*}
Since $H$ is linear $\hat{\xi} = H(\hat{x}) = H(x) + H(\nu) := \xi + V$ thus
\begin{equation*}
 Y_m = \hat{X} (I+ \xi + V)  = \hat{X} (\delta X + V) = X +  \hat{X} V
\end{equation*}
Note that $XV=  \hat{X} \delta X V=  \hat{X} (I+\xi)  V =  \hat{X} V +   \hat{X} \xi V \approx \hat{X} V$ since $\xi V$ is a second-order term, and so the
pose output from odometry-aided measured scan matching is
\begin{equation}
 Y_m = X + XV, \qquad V=H(\nu)=\begin{bmatrix} S(\nu_R) & \nu_p \\ 0 & 0\end{bmatrix}
\label{eq:finalYmICPb}
\end{equation}
where $\text{cov}(\nu)=C$ is given by~\eqref{eq:covICPb}.

\section{Estimator Design}

\subsection{Invariant Observer}
\label{sec:invobs}

We first design an invariant observer for the noise-free system~\eqref{eq:dynSE3},~\eqref{eq:outSE3nonoise} by following the method in~\cite{BMR08,BMR09}; a
tutorial presentation is available in~\cite{BL13}. Let $G$ be a Lie group acting on the system dynamics $\dot{x}=f(x,u)$ state and input spaces via the Lie
group actions $\varphi_g:G\times x \to x$ and $\psi_g:G\times u \to u$, respectively. This system is termed $G$-invariant if
\begin{equation*}
\frac{d}{dt} \varphi_g(x) = f(\varphi_g(x),\psi_g(u)), \qquad \forall g\in G
\end{equation*}
Finding the actions making the system invariant is non-systematic, although it is based on the physics of the problem. For dynamics~\eqref{eq:dynSE3},
choosing $G=SE(3)$ and $\varphi_g x = gx$, $\psi_g u =u$ can be directly verified to provide $G$-invariance, which we refer to as left
invariance since $\varphi_g=L_g$. Left invariance physically represents applying a constant rigid-body transformation to ground-fixed frame vector coordinates,
for which
the governing dynamics~\eqref{eq:dynSE3} still hold. The property of $G$-invariance is not unique, for instance $G=SE(3)$ with $\varphi_g
x=x g$, $\psi_g u = g^{-1} u g$ verifies another $G$-invariance of~\eqref{eq:dynSE3} known as right, which was employed in~\cite{Bon12,HBG12}. However we
will be employing 
left
invariance for reasons explained in Section~\ref{sec:IEKF}.

The remaining steps for obtaining an invariant observer are systematic. The actions $\varphi_g$, $\psi_g$ induce a Lie group action on the output space
$\rho_g:G
\times y \to y$ satisfying the $G$-equivariance
\begin{equation*}
\rho_g(y) = h(\varphi_g(x),\psi_g(u)), \qquad \forall g\in G
\end{equation*}
where in our case~\eqref{eq:outSE3nonoise} with left invariance gives $\rho_g y=g y$. The subsequent design steps consist of computing the moving frame,
finding the
complete set of invariants, obtaining the invariant output error, invariant estimation error and invariant frame, and obtaining the structure of the invariant
observer as well as its associated invariant estimation error dynamics. The details of the steps are fully discussed in~\cite{BMR08,BL13} and will not be
reprinted here due to space constraints. For the present system~\eqref{eq:dynSE3},~\eqref{eq:outSE3nonoise} under left invariance, we obtain the moving frame
$\gamma(x)=X$; the complete set of invariants
\begin{equation*}
I(x,u) =\psi_{\gamma(x)}(u)=\Omega, \quad
J_h(x,y) = \rho_{\gamma(x)}(y)=X^{-1} Y
\end{equation*}
the invariant output error
 \begin{equation*}
 E(\hat{x},u,y)=J_h(\hat{x},h(\hat{x},u)) - J_h(\hat{x},y) = I-\hat{X}^{-1} Y
 \end{equation*}
the invariant estimation error
 \begin{equation*}
\eta(x,\hat{x})=\varphi_{\gamma(x)}(\hat{x})-\varphi_{\gamma(x)}(x) = X^{-1} \hat{X}-I
 \end{equation*}
 redefined as $\eta = X^{-1} \hat{X}$ for convenience; the invariant frame
\begin{equation*}
w_i^R = d \varphi_x v_i^R
=   \begin{bmatrix}
R S(e_i)  & 0 \\
0 & 0
\end{bmatrix},\,  
w_i^p = d \varphi_x  v_i^p
=   \begin{bmatrix}
0  & R e_i \\
0 & 0
\end{bmatrix}
\end{equation*}
and the invariant observer
\begin{align}
 \dot{\hat{X}} &= \hat{X} \Omega + \sum_{i=1}^3 \mathcal{L}_i^R \begin{bmatrix}
\hat{R} S(e_i)  & 0 \\
0 & 0
\end{bmatrix}  + \sum_{i=1}^3 \mathcal{L}_i^p \begin{bmatrix}
0  & \hat{R} e_i \\
0 & 0
\end{bmatrix}  \nonumber \\
&= \hat{X} \Omega + \begin{bmatrix}
\hat{R}   & \hat{p} \\
0 & 1
\end{bmatrix}   \begin{bmatrix}
 S(\mathcal{L}^R)  & \mathcal{L}^p \\
0 & 0
\end{bmatrix} := \hat{X} \Omega + \hat{X} L \label{eq:invobs}
\end{align}
where $\mathcal{L}^R\in\mathbb{R}^3$, $\mathcal{L}^p\in\mathbb{R}^3$ are smooth functions of $I(\hat{x},u)$, $E(\hat{x},u,y)$ such that
$se(3) \ni L(I,0)=0$.

We cannot directly employ $E= I-\hat{X}^{-1} Y$ to form the gain term $L$ in~\eqref{eq:invobs} since
$I-\hat{X}^{-1}Y \notin SE(3)$. Instead we take $L=\pi(\hat{X}^{-1} Y)$ since
by~\eqref{eq:XminusIapprox}, $\hat{X}^{-1} Y = I \Longleftrightarrow L=0$ as required. We add the observer gain matrix $K \in \mathbb{R}^{6\times 6}$ as
\begin{equation*}
L = H\circ K \circ H^{-1} [ \pi(\hat{X}^{-1} Y) ]
\end{equation*}
With this $L$ the dynamics of $\eta= X^{-1} \hat{X}$ compute to
\begin{equation}
\begin{aligned}
 \dot{\eta} &= X^{-1} X \Omega X^{-1} \hat{X} + X^{-1} \hat{X} \Omega + X^{-1} \hat{X} L \\
 &= \Omega \eta + \eta \Omega + \eta H\circ K\circ H^{-1}[ \pi ( \eta^{-1})]
\end{aligned}
\label{eq:ieed}
\end{equation}
and stabilizing the (nonlinear) dynamics~\eqref{eq:ieed} to $\overline{\eta}=I$ by choice of gains $K$ leads to an asymptotically stable nonlinear
observer~\eqref{eq:invobs}. The stabilization process is simplified by~\eqref{eq:ieed} not being dependent on the estimated state $\hat{X}$; indeed the
fundamental
feature of the invariant observer is that it guarantees $\dot{\eta} =
\Upsilon(\eta,I(\hat{x},u))$~\cite[Theorem~2]{BMR08} with $I(\hat{x},u)=\Omega$ in the present example, which simplifies gain selection over the general case,
but does not make it systematic. For this reason we will use the Invariant EKF method to obtain stabilizing gains.

\subsection{Invariant EKF}
\label{sec:IEKF}

We first recall the continuous-time EKF algorithm. Given the nonlinear system $\dot{x}=f(x,u,w)$, $y=h(x,v)$ where $w$ and $v$ are the process and output
Gaussian noise vectors, we linearize about $(x,u,w,y,v)=(\hat{x},u,0,\hat{y},0)$, a
nominal (noise-free) trajectory of the system,
$\dot{\hat{x}}=f(\hat{x},u,0)$, $\hat{y}=h(\hat{x},0)$, and obtain
\begin{equation*}
\begin{aligned}
\delta \dot{x}&=A \delta x + B w \\
\delta y&=C \delta x + D v
\end{aligned}
\end{equation*}
The Kalman Filter for this linearized time-varying system is
\begin{equation}
\begin{aligned}
\delta\dot{\hat{x}}&= A \delta \hat{x} + K(\delta y-C \delta \hat{x}) \\
K&=PC^T(DRD^T)^{-1} \\
\dot{P}&= A P + P A^T - PC^T(DRD^T)^{-1}CP + BQB^T
\end{aligned}
\label{eq:ctEKF}
\end{equation}
and $\delta \hat{x} +\hat{x}$ then becomes the estimated state of the original nonlinear system. The above EKF possesses the estimation error
$\varepsilon=\delta \hat{x}-\delta x$ dynamics
\begin{equation*}
\dot{\varepsilon} = (A-KC)\varepsilon - Bw + KDv
\end{equation*}

The Invariant EKF is a systematic approach to computing the gains $K$ of an invariant observer by linearizing
its
invariant estimation error dynamics. We first introduce input and output noise terms
$\tilde{u}=u+w$, $\tilde{y}=y+v$ such that $w$ and $v$ preserve the $G$-invariance of the system as
$\dot{x}=f(x,\tilde{u}-w)$ and $\dot{\hat{x}}=F(\hat{x},\tilde{u},y+v)$. Using these we compute $\dot{\eta}=\Upsilon(\eta,I(\hat{x},u),w,v)$ then linearize
about $\eta=\overline{\eta}$, $w=v=0$ to obtain the form
\begin{equation*}
\delta \dot{\eta} = (A-KC) \delta \eta - B w + K D v
\end{equation*}
We then read off the $(A,B,C,D)$ matrices and employ the conventional EKF formulas~\eqref{eq:ctEKF} to compute stabilizing gains $K$ for the invariant observer.
The
interest of the Invariant EKF is the reduced dependence of the linearized system on the estimated system trajectory of the original system, specifically
only through the latter's estimated invariants $I(\hat{x},u)$. In our present case $I(\hat{x},u)=\Omega$ thus the $(A,B,C,D)$ matrices are
guaranteed not to
depend on the estimated state, which increases the filter's robustness to poor state estimates and precludes divergence (c.f.~Section~\ref{sec:intro}).

Returning to~\eqref{eq:dynSE3},~\eqref{eq:outSE3nonoise}, we first need to introduce $w$ and $v$ noise terms which preserve $G$-invariance and $G$-equivariance.
As discussed in
Section~\ref{sec:invobs} the left invariance case corresponds to transforming ground-fixed frame vector coordinates, and so introducing noise terms expressed in
the body-fixed frame will not affect the invariance of the system. For inertial sensors we write
\begin{equation*}
\Omega_m = \begin{bmatrix}
S(\omega ) & \mu   \\
0 & 0
\end{bmatrix} + \begin{bmatrix}
S( \nu_\omega) & \nu_\mu  \\
0 & 0
\end{bmatrix}:= \Omega + W
\end{equation*}
i.e.~use an additive sensor noise model, which is standard in inertial navigation design~\cite{Farrell} since $\text{cov}(W)$
can be directly identified from logged sensor data. This is precisely the reason why we chose to use the left-invariant version of the observer in
Section~\ref{sec:invobs}. For the output equation we have obtained~\eqref{eq:finalYmICPb} where $V$ represents body-frame noise
terms due to the ICP alignment being performed in the body-fixed frame.

Introducing $w$ and $v$ as $\dot{x}=f(x,\tilde{u}-w)$,
$\dot{\hat{x}}=F(\hat{x},\tilde{u},y+v)$ into dynamics~\eqref{eq:dynSE3} and observer~\eqref{eq:invobs} we have
\begin{equation*}
\begin{aligned}
\dot{X}&=X ( \Omega_m - W ) \\
\dot{\hat{X}} &= \hat{X} \Omega_m + \hat{X} L_m, \quad L_m:=H\circ K \circ H^{-1} [ \pi(\hat{X}^{-1} Y_m) ]
\end{aligned}
\end{equation*}
and computing $\dot{\eta}$ where $\eta= X^{-1} \hat{X}$ gives
 \begin{equation*}
 \begin{aligned}
 \dot{\eta} &= - X^{-1} X(\Omega_m - W)  X^{-1} \hat{X} + X^{-1} (\hat{X} \Omega_m + \hat{X} L_m ) \\
  &= \eta \Omega_m -\Omega_m \eta  + W \eta + \eta L_m
 \end{aligned}
 \end{equation*}
Linearizing the above around $\eta=\overline{\eta}$, $w=v=0$ we have 
$\eta \approx I + \xi$, $\eta^{-1} \approx I -
\xi$ by \eqref{eq:Xapprox},~\eqref{eq:Xinvapprox} and so
\begin{equation*}
 \begin{aligned}
\dot{\xi} &= (I+\xi)\Omega_m - \Omega_m(I+\xi) + W (I+\xi) + (I+\xi) L_m \\
&= \xi \Omega_m - \Omega_m \xi + W + H\circ K \circ H^{-1} [ \pi(\hat{X}^{-1} Y_m) ]
\end{aligned}
\end{equation*}
and
\begin{equation*}
 \begin{aligned}
\pi(\hat{X}^{-1} Y_m) &= \hat{X}^{-1} Y_m - I = \eta^{-1} (I+V) - I \\
&= (I-\xi)(I+V) - I = V- \xi
 \end{aligned}
\end{equation*}
By~\eqref{eq:se3algebraform} we define $\xi:=H([\zeta_R \quad \zeta_p])$, $\Omega_m:=H([\omega_m \quad \mu_m])$,
$W:=H([\nu_\omega \quad \nu_\mu])$, $V:=H([\nu_R \quad \nu_p])$ and write $\dot{\xi}$ as
\begin{equation*}
\begin{aligned}
\begin{bmatrix}
\dot{\zeta}_R \\ \dot{\zeta}_p
\end{bmatrix}
&= \begin{bmatrix}
-S(\omega_m) & 0 \\ -S(\mu_m) & -S(\omega_m)
\end{bmatrix} \begin{bmatrix}
\zeta_R \\ \zeta_p
\end{bmatrix} + \begin{bmatrix}
\nu_\omega
\\
 \nu_\mu
\end{bmatrix} \\&+ K \left( \begin{bmatrix} \nu_R \\ \nu_p \end{bmatrix} - \begin{bmatrix}
\zeta_R \\ \zeta_p
\end{bmatrix} \right)
\end{aligned}
 \end{equation*}
By matching the above with $\delta \dot{\eta} = (A-KC) \delta \eta - B w + K D v$ we read off
\begin{equation*}
\begin{aligned}
 A&= \begin{bmatrix}
-S(\omega_m) & 0 \\ -S(\mu_m) & -S(\omega_m)
\end{bmatrix}, \quad B= \begin{bmatrix}
-I & 0 \\ 0 & -I
\end{bmatrix}\\
C&= \begin{bmatrix}
I & 0 \\ 0 & I
\end{bmatrix}, \quad D= \begin{bmatrix}
I & 0 \\ 0 & I
\end{bmatrix}
\end{aligned}
\end{equation*}
Just as predicted, the linearized system matrices do not depend on the estimated trajectory $\hat{X}$ but only on $\Omega=I(\hat{x},u)$. Using $(A,B,C,D)$
with~\eqref{eq:ctEKF} we compute the gain $K$ of the invariant observer by
\begin{equation*}
\begin{aligned}
\dot{P}&=AP + PA^T-PV^{-1}P + W \\
K&=P V^{-1}
\end{aligned}
\end{equation*}

\section{Experimental Validation}
\label{sec:expvalidation}

\subsection{Hardware Platform}

\begin{figure}[thpb]
    \center
     \includegraphics[width=0.4\columnwidth]{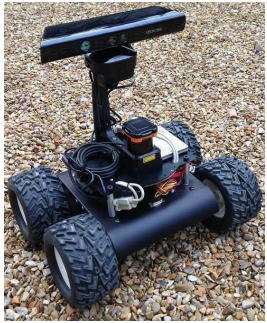}
  \caption{The Wifibot Lab v4 Robot}
 \label{fig:wifibotpic}
\end{figure}

The wheeled robot used for our experiments is shown in Figure~\ref{fig:wifibotpic}. The robot is equipped with an Intel Core i5-based
single-board computer, WLAN 802.11g wireless networking, all-wheel drive via 12V brushless DC motors, and a Kinect camera providing 3-D point
cloud scans of the environment. The odometry data at $50\text{ Hz}$ and Kinect point clouds at
$5\text{ Hz}$ are passed to the IEKF which estimates the plane position $(x,y)$ and heading angle $\psi$ of the vehicle. The robot is also equipped with a
Hokuyo UTM-30LX
LiDAR which provides centimeter-level positioning accuracy through proprietary SLAM code~\cite{SBTB11}. This LiDAR-derived data is used solely to provide
a reference trajectory when plotting results.

\subsection{Experimental Results}

\subsubsection{Straight line trajectory}

\begin{figure}[thpb]
    \center
     \includegraphics[width=0.4\columnwidth]{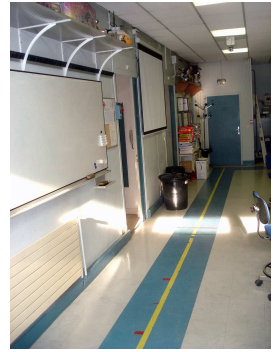}
  \caption{First experiment field}
 \label{fig:labo}
\end{figure}

\begin{figure}[thpb]
    \center
     \includegraphics[width=0.67\columnwidth]{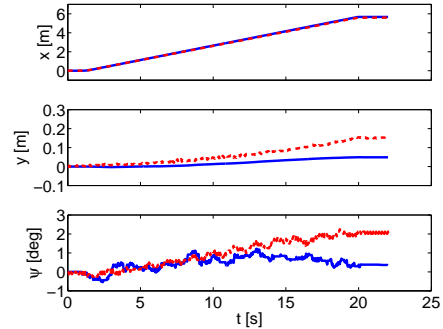}
  \caption{Estimates for linear trajectory: IEKF (solid), reference (dashed)}
 \label{fig:exp1all}
\end{figure}

The first experiment field is the laboratory shown in Figure~\ref{fig:labo}. The robot begins stationary at the bottom edge of the picture then advances
with constant velocity in a straight line along the marking on the floor towards the far door, where it comes to a stop. The motion was commanded
in open-loop and the robot's
trajectory exhibited a slight veer to the left over the full length. The state estimates by the IEKF are plotted
against the LiDAR-derived reference states in Figure~\ref{fig:exp1all}.

Figure~\ref{fig:exp1all} illustrates that the IEKF provides coherent estimates: both the forward motion and the left veer are correctly rendered in the
estimates. The RMS discrepancy of the estimates from the reference trajectory is $(4.5,5.3)\text{ cm}$ for the $(x,y)$ positions and $0.9^\circ$ for the heading
angle $\psi$.

\subsubsection{Circular trajectory}

\begin{figure}[thpb]
    \center
     \includegraphics[width=0.6\columnwidth]{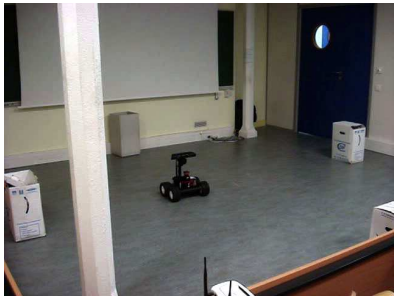}
  \caption{Second experiment field}
 \label{fig:circlepic}
\end{figure}

\begin{figure}[thpb]
    \center
     \includegraphics[width=0.67\columnwidth]{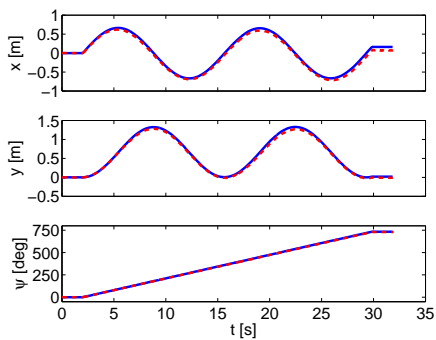}
  \caption{Estimates for circular trajectory: IEKF (solid), reference (dashed)}
 \label{fig:exp2overhead}
\end{figure}

The second experiment consists of executing a circular trajectory in the environment shown in
Figure~\ref{fig:circlepic}. The test area was surrounded on all sides by a bounding wall,
and a number of visual landmarks were placed around the test area to provide better scan matching conditions during turning
maneuvers. The robot began
stationary, executed two concentric counter-clockwise circles with a constant velocity, and then stopped at its initial position. The overhead
position estimates from the IEKF are plotted against
the LiDAR-derived reference trajectory in Figure~\ref{fig:exp2overhead}.

The IEKF estimates once again follow the LiDAR reference trajectory; the computed RMS errors for $(x,y)$ and $\psi$ are $(5.1, 3.5)\text{ cm}$ and $2.5^\circ$,
respectively.

\section{Conclusions}

We have described a scan matching SLAM design based on an Invariant EKF. The proposed approach guarantees robustness of the filter to poor state
estimates $\hat{X}$, which may lead to degraded performance or even destabilize the filter in the conventional EKF case. The Invariant EKF was successfully
implemented in hardware and performed well in
experiments, making it a
promising candidate for more complex SLAM applications such as mobile outdoor cartography.

\section{Acknowledgements}

We thank Tony No{\"e}l for his extensive help with setting up the Wifibot platform and running hardware experiments. The work reported in this paper was
partly supported by the Cap Digital Business Cluster TerraMobilita Project.


\bibliographystyle{IEEEtran}
\bibliography{IEEEabrv,bibliography}

\end{document}